\documentclass[pss]{wiley2sp} 
\usepackage{amsmath}
\usepackage{float}

\begin{document}

\title{Symmetry breaking and low energy conformational fluctuations in amorphous graphene}

\titlerunning{Symmetry breaking and conformational fluctuations in $\alpha$-G }

\author{%
  Y.\ Li\textsuperscript{\textsf{\bfseries 1}},
  D.\ A.\ Drabold\textsuperscript{\Ast,\textsf{\bfseries 1}}}

\authorrunning{Y.\ Li and D.\ A.\ Drabold.}

\mail{e-mail
  \textsf{drabold@ohio.edu}, Phone:
  +1 740 593 1715, Fax: +1 740 593 0433}

\institute{%
  \textsuperscript{1}\, Department of Physics and Astronomy, Ohio University, Athens, OH 45701, USA}

\received{XXXX, revised XXXX, accepted XXXX} 
\published{XXXX} 


\abstract{%
\abstcol{%
Recently, the prospects for amorphous phases of graphene ($\alpha$-G) have been explored computationally. Initial models were flat, and contained odd-member rings, while maintaining three-fold coordination and $sp^2$ bonding. Upon relaxation, puckering occurs, and may be traced to the existence of pentagons, in analogy with the situation for fullerenes. In this work, we systematically explore the inherent structures with energy close to the flat starting structure. As expected, the planar symmetry can be broken in various ways, which we characterize for 800-atom model of  $\alpha$-G, always using local basis
  }{%
 density functional techniques. The classical normal modes of various structural models are discussed, with an emphasis on imaginary modes indicating the evolution from flat to puckered. We also discuss very low energy conformational fluctuations akin to those seen previously in amorphous silicon \cite{md_drabold}, and reflect on the nature of the amorphous ``ground state'' within a network of fixed topology. For completeness, high energy modes were also computed, and are found to be associated with strained parts of the network.}}

\maketitle   

\section{Introduction}
\label{sec:intro}

A key feature of matter in a disordered state is the existence of large numbers of conformations with essentially degenerate energies, which may also be accessible with small energy cost. This is in contrast with crystals, which possess long range order, few polymorphs and typically a deep energy minimum and large energy barrier. The ground state is thus sharply defined, and the only low energy excitations are phonons -- transitions to other structures are prohibited. Where realistic models of disordered systems are concerned,  few attempts have been made to quantitatively characterize the number, energetics and proximity (in the sense of barrier) of these states. In his inherent structure formulation of statistical mechanics, Stillinger argued that the number of minima scales like $N!exp(\alpha N)$\cite{scn_potential}, where $N$ is the number of atoms in the model, and $\alpha$ is a positive system-dependent constant. $\alpha$ was estimated to be around 0.8 in a monatomic liquid\cite{alpha_mona}, and flexible organic molecules exhibit larger $\alpha$, as in the fragile glass former ortho-terphenyl where $\alpha$ is around 13.14\cite{alpha_otp}. For temperatures well below the melting point, these local energy minima are denoted inherent structures. Each inherent structure is confined to its own basin, which refers to a set of configurations can be mapped to the inherent structure by potential energy minimization\cite{scn_potential}.

To further motivate this work, consider the following \textit{gedanken} experiment. Consider a sequence of molecules with $N$ atoms. It is well known that as $N$ increases, the number of minima accessible to the molecule also increase, and such conformations are extensively studied in chemistry\cite{trans_still}\cite{inherent_still}\cite{alpha_value}\cite{saddle_doye}. While it is unlikely that a rigorous theory quantifying these minima as a function of $N$ can be formulated, it is clear from computer experiments that the number of minima grow drastically with $N$. For most molecular systems it is difficult to be certain that the global minimum structure has been found in a simulation as there are so many metastable minima in which the system can be trapped. In this paper, we are concerned with the even more intractable problem of characterizing the minima, or potential energy surface of a disordered condensed matter system in both two-dimensional (2D) and three-dimensional (3D) cases. For a 3D crystal, if one introduces small random distortions and relaxes the disturbed system, it returns to exactly the same structure. This is \textit{not} true for disordered systems. For 3D, we find there exits a continuum of metastable minima for $\alpha$-Si, in which a number of tiny distortions of bond angles (and to a much lesser extend bond lengths) yield a distinct energy-degenerate conformation, which reveals the existence of an extraordinarily flat potential energy landscape (PEL). Fedders and Drabold showed that starting from a well-relaxed \textit{a}-Si:H model, quenching a set of ``snapshots'' during a constant-T MD simulation never returns to the exact initial state, instead they fall into minima that are topologically equivalent (eg.\ with the same network connectivity), but with small variations in bond angles and bond lengths\cite{md_drabold}. We find that the behavior in $\alpha$-Si is consistent with $\alpha$-Si:H. The case of $\alpha$-G is different in the following sense. Like $\alpha$-Si, we find a continuum of essentially energy degenerate states in which minute (but ``real'') variations in bond angles and bond lengths are displayed. Of course, such states retain identical connectivity since their energies are identical to within a few micro eV. However, $\alpha$-G also exhibits a variety of local energy minima associated with different puckering. These structures usually have similar energies (within $\sim 10 meV$), but a significant barrier separating them. Thus, the picture that emerges of the $\alpha$-G energy landscape is a variety of inherent structures (with varying puckering) with slightly varying energies but substantial barriers between them, and in each of these basins is associated with a particular puckered state, an ambiguously defined minimum with small variations in bond angles and bond lengths accessible as we describe in detail below in Sec.\ \ref{sec:confor_fluc}.

Beside the work described above on ultra-low energy excitations, we also discuss other phenomena peculiar to $\alpha$-G. There has been intensive study in understanding the properties of crystalline graphene, but little is securely understood about amorphous phases. Recent electron bombardment experiments have revealed the existence of amorphous graphene\cite{kota_prb} \cite{kota_prl}. Clear images of regions of amorphous graphene have been published by Meyer\cite{image_ag}. In previous work, we observed that  planar amorphous graphene is extremely sensitive to out-of-plane distortions \cite{grap_pucker}. Similar behavior has been verified in amorphous graphene by experiment and other calculations \cite{lusk_prl} \cite{metal_prb} and also in silicon nanosheets \cite{si_pucker}. We have found different initial conditions in breaking the planar symmetry of amorphous graphene model lead to distinct puckered states after relaxation \cite{grap_pucker}. These states exhibit little difference in topological properties, ie.\ ring statistics and coordination number. However, the total energy differences between these metastable states are around $0.02 eV$ per atom, and the full width of the puckering along the original normal direction is around $6-7 \mathring{A}$. 

\begin{figure}[htb]
	\centering
	\subfloat[Top view of 800-atom crystalline graphene.]{%
		\includegraphics[scale=0.38]{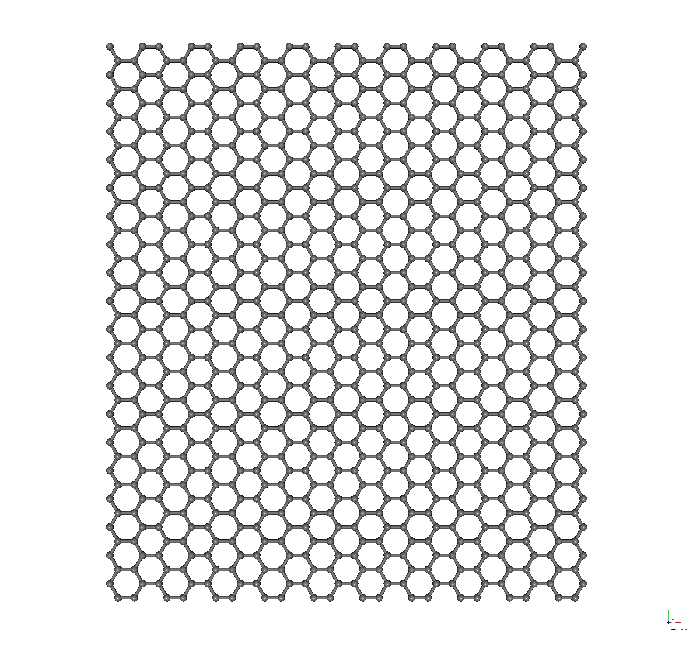}
		\label{fig:800c-g}
}\\
	\subfloat[Top view of 800-atom amorphous graphene due to He and Thorpe relaxed by us.]{%
		\includegraphics[scale=0.38]{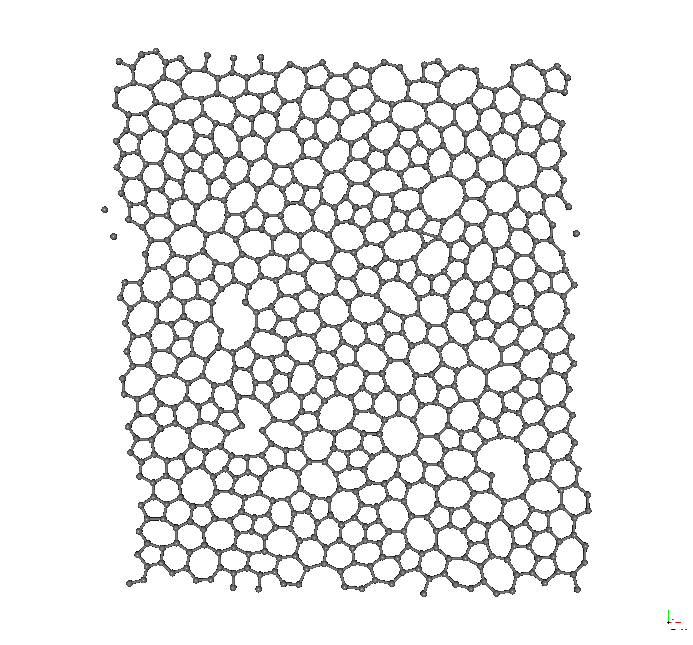}
		\label{fig:800a-g}
}
	\caption{Comparison between crystalline and amorphous phases of graphene.}
	\label{fig:com_graphene}
\end{figure}

A natural complement to these studies is an exploration of low-frequency classical normal modes. These modes turn out to be rather delocalized. As presented for two level systems, the tunneling between two equilibrium states triggers a number of low-energy excitations\cite{tls_anderson}\cite{tls_phillips}. Details are discussed in Sec.\ \ref{sec:modes}. In Sec.\ \ref{sec:conclusion}, we summarize our findings about the degenerate states and localized imaginary-, low- and high-frequency vibrational modes of amorphous graphene.

\section{Models}
\label{sec:mol}

To approach this problem computationally, we employ an 800-atom amorphous graphene model (800 $\alpha$-g) due to He and Thorpe generated by introducing Stone-Wales defects into a perfect honeycomb lattice and a WWW annealing scheme\cite{www_paper}. This model has perfect threefold coordination with varying concentration of 5, 6 and 7 member rings\cite{grap_drabold} and is a practical realization of the continuous random network (CRN) concept proposed by Zachariason\cite{crn_zach}. The comparison between crystalline and amorphous graphene is shown in Fig.\ \ref{fig:com_graphene}. We have relaxed the 800-atom $\alpha$-g model with an accurate \textit{ab-initio} code, while small rearrangements occurred and planar symmetry was preserved. We found that by very slightly breaking the planar symmetry (by randomly moving each atom by $\sim0.01 \mathring{A}$) and performing a structural optimization, the resulting minima were always puckered\cite{grap_pucker}.

The amorphous Si model we employ is a realistic 64-atom model (64 $\alpha$-Si), generated by Barkema and Mousseau using a modified form of WWW algorithm\cite{64si_prep}. This model has perfect fourfold coordination and within the limitations of its small size, to our knowledge is not in significant contradiction to any experiment.

\section{Procedure}
\label{sec:md}

Our calculations  are performed with an \textit{ab-initio} program SIESTA\cite{siesta}, using pseudopotentials and the Perdew-Zunger parameterization of the local-density approximation (LDA) with a single-$\zeta$ basis and Harris-Functional at a constant volume. For earlier simulations the details of method are described in \cite{drabold8} and \cite{drabold37}. To investigate the nature of minima on the potential energy surface, we employ a method proposed by Fedders and Drabold\cite{md_drabold} similar to the conformational space annealing approach mentioned in \cite{pel_wales}, which has been used in locating and predicting low-energy conformations of various proteins\cite{csa_1}\cite{csa_2}\cite{csa_3}\cite{csa_4}. First, starting with a perfectly relaxed model (in our case 800 $\alpha$-g and 64 $\alpha$-Si), we run a sequence of four parallel simulations. We let the network evolve for $8.0 ps$ at four different mean temperatures of 20K, 500K, 600K or 900K. The target temperatures are achieved by velocity rescaling. From these simulations,we drew instantaneous snapshots of the configuration by saving the instantaneous configuration every $0.15 ps$, then rapidly quench these to 0K and let the quenched configurations evolve around 0K for a few $ps$ to find the metastable minimum (or inherent structure) associated with the initial snapshots.

To investigate the topological changes between these minima (quenched configurations from snapshots), we use two autocorrelation functions as defined by Fedders and Drabold\cite{md_drabold}:
\begin{equation}
\label{eq:md_auto1}
\Delta \theta(t_1,t_2)=\displaystyle\sum_i((\theta_i(t_1)-\theta_i(t_2))^2/N)^{\frac{1}{2}}
\end{equation}
and
\begin{equation}
\label{eq:md_auto2}
\Delta r(t_1,t_2)=\displaystyle\sum_i((r_i(t_1)-r_i(t_2))^2/N)^{\frac{1}{2}}
\end{equation}
In Eq.\ \ref{eq:md_auto1} the index $i$ runs through all the bond angles where $\theta_i$ is the $i^{th}$ bond angle. In Eq.\ \ref{eq:md_auto2} the index $i$ runs over all nearest-neighbor pairs where $r_i$ is the $i^{th}$ is the distance between a pair. All the atoms in quenched supercells are threefold, making the definitions of $\theta_i$ and $r_i$ well defined. The times $t_1$ and $t_2$ refer to the quenched  snapshots. These autocorrelation functions provide a close view of how thermal MD simulations induce transitions between different energy basins.

\section{Discussion}
\label{sec:discuss}

We break the discussion into three parts. First, the nature of pentagonal puckering, conformation fluctuations and finally an analysis of the classical vibrational modes.

\subsection{Symmetry Breaking}

\begin{figure}[htb]
\includegraphics[scale=0.15]{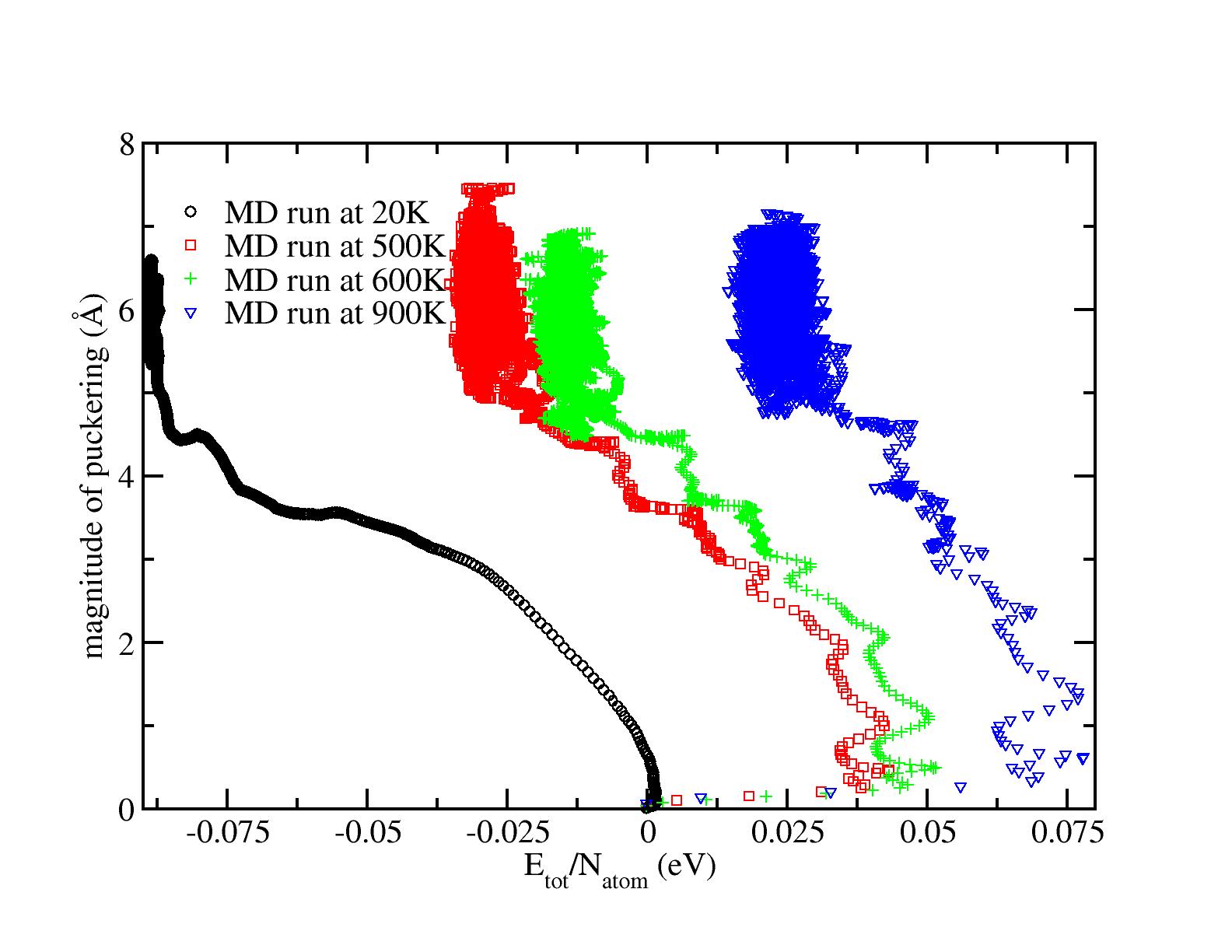}
\caption{Color. Correlation between the total energy per atom and magnitude of puckering for constant temperature MD simulations. The zero total energy refers to the total energy of original flat 800 $\alpha$-g model.}
\label{fig:md_zdiff}
\end{figure}

As we have shown in \cite{grap_pucker}, the original flat 800-atom $\alpha$-g model is exceedingly sensitive to transverse distortion, and then loses planar symmetry lowering the total energy of the supercell. In every case, even at $T=20K$, the planar symmetry breaks and the system puckers: thermal disorder is sufficient to induce puckering. Fig.\ \ref{fig:md_zdiff} shows the relation between the total energy of the system and maximum separation of atoms along the normal direction (magnitude of puckering) in constant-T MD simulations at the indicated temperatures. In the language of PEL, starting from the flat 800 $\alpha$-g, four MD simulations overcome tiny energy barriers and take a down-hill path to regions with lower energy. Thus the flat 800 $\alpha$-g model can be considered as an exceptionally shallow basin  on the PEL. The barrier to puckering from flat is a few micro eV for this Hamiltonian.

\subsection{Conformational Fluctuations}
\label{sec:confor_fluc}

\begin{table*}[h]
	\centering
	\caption{Average value and standard deviation of $E_{tot}/N_{atom}$, $\Delta r(t_1,t_2)$ and $\Delta \theta(t_1,t_2)$ of quenched configurations from MD runs in the time period from 3.6 to 8.0 ps, where $t_1=1.05 ps$.}
	\begin{tabular}[htbp]{@{}lllllll@{}}
		\hline
		T(K) & $\overline{E_{tot}/N_{atom}}$ (eV) & $\sigma_{E_{tot}/N_{atom}}$ (eV) & $\overline{\Delta r(1,t_2)}$ ($\mathring{A}$) & $\sigma_{\Delta r}$ ($\mathring{A}$) & $\overline{\Delta \theta(1,t_2)}$ ($\circ$) & $\sigma_{\Delta \theta}$ ($\circ$) \\
		\hline
		500K & $-9.595\times10^{-2}$ & $8.855\times10^{-5}$ & $4.907\times10{-2}$ & $2.551\times10^{-4}$ & 3.736 & $1.814\times10^{-2}$ \\
		600K & $-9.435\times10^{-2}$ & $9.917\times10^{-5}$ & $7.503\times10^{-2}$ & $1.973\times10^{-4}$ & 4.181 & $6.464\times10^{-2}$ \\
		900K & $-9.716\times10^{-2}$ & $2.091\times10^{-4}$ & $2.516\times10^{-2}$ & $8.08\times10^{-3}$ & 1.725 & 0.146 \\
		\hline
	\end{tabular}
	\label{tab:info_para}
\end{table*}

\begin{figure}[htb]
\includegraphics[scale=0.15]{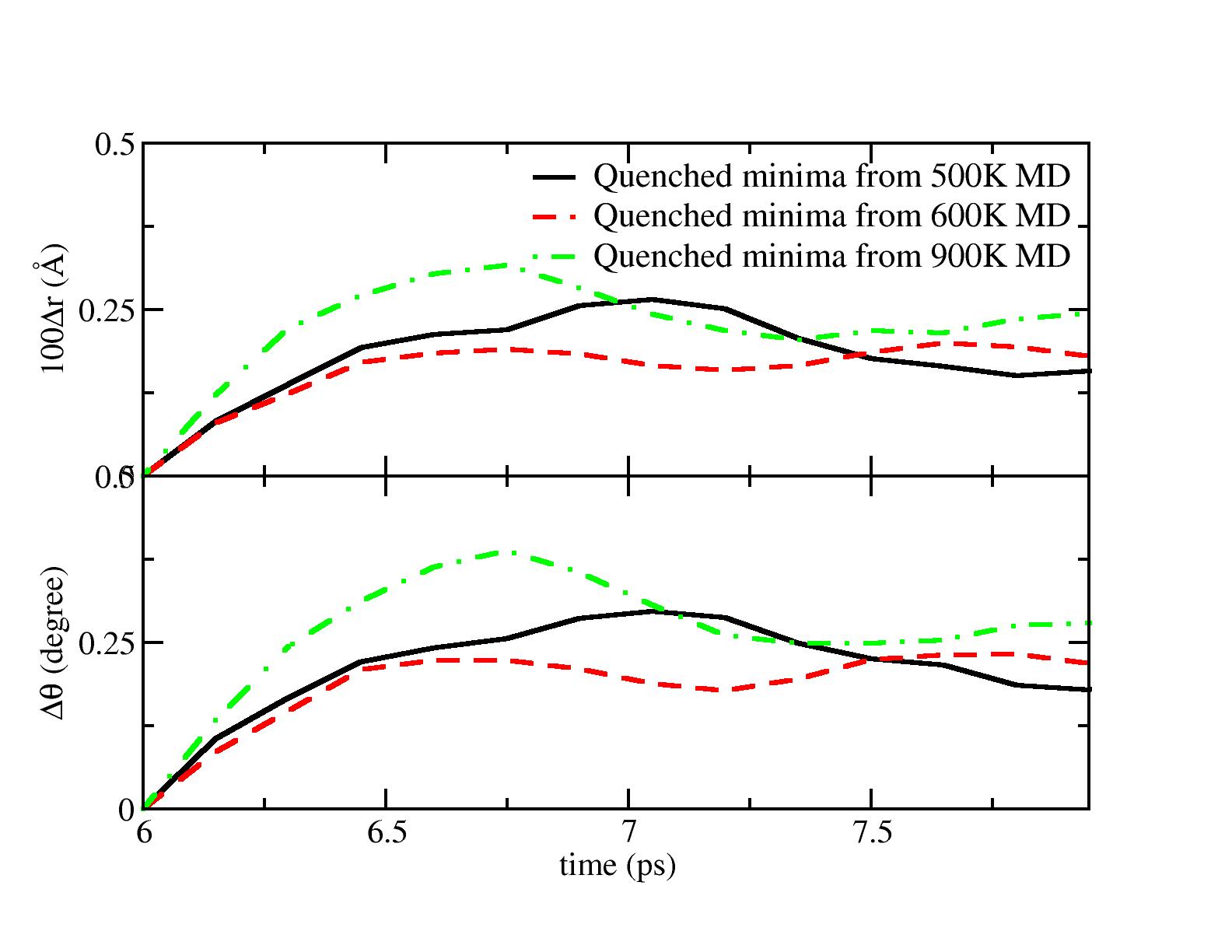}
\caption{Color. Time variation of two autocorrelation functions. This figure shows autocorrelation functions of $\Delta \theta(t_1,t_2)$ and $100\Delta r(t_1,t_2)$ for $t_1=6.0 ps$ and $t_2$ varying from $6.0$ to $7.95 ps$. The temperatures are 500K, 600K and 900K. The functions appear to be continuous.}
\label{fig:md_func}
\end{figure}
        
\begin{figure}[htb]
\includegraphics[scale=0.2]{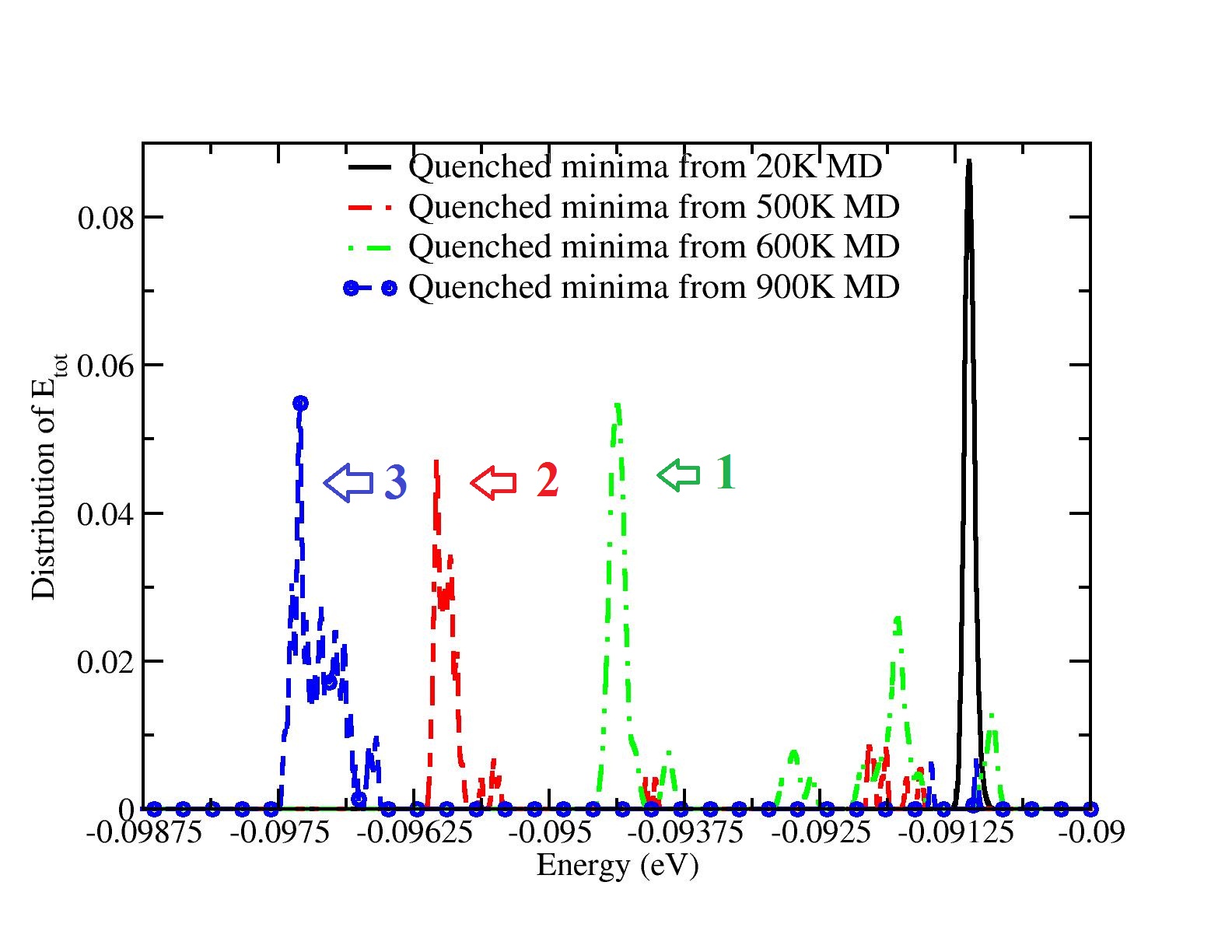}
\caption{Color. Total energy distribution functions of quenched supercells from MD runs under 20K, 500K, 600K and 900K. The total energy of original flat 800 a-g is considered as $0eV$. Distinct structures correspond to different puckered states, broadening within each major peak from conformational variations. Three major peaks from MD runs at 600K, 500K and 900K are labeled as 1, 2 and 3 respectively.}
\label{fig:md_diseng}
\end{figure}
    
The quenching procedures at sequential timesteps yield basins on the PEL. Here we show the calculations at average temperatures of 500K, 600K and 900K. These results of MD runs at different temperatures exhibit consistency with each other. Fig.\ \ref{fig:md_func} shows how these two autocorrelation functions vary with time of snapshots. Since temperatures of all the MD simulations achieve equilibrium after 6.0 ps, here the autocorrelation functions are calculated with $t_1=6.0 ps$. It appears $\Delta \theta(t_1,t_2)$ and $\Delta r(t_1,t_2)$ from three MD runs at different temperatures are qualitatively similar. They increase linearly and approach a constant. The continuity of the curves in Fig.\ \ref{fig:md_func} suggests that there is a continuum of states, accessible albeit structurally varying only in very modest ways.

\begin{figure}[htb]
\includegraphics[scale=0.3]{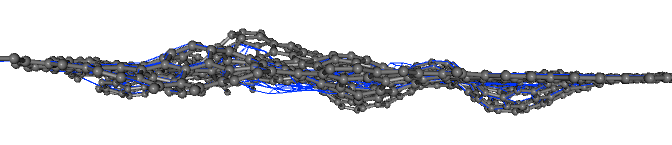}
\caption{Color. Side view of two quenched configurations. Gray balls and sticks show the configuration from 900K MD, and blue lines represent the one from 500K.}
\label{fig:eng_side}
\end{figure}

The total energy distributions of all the quenched supercells from MD runs at different temperatures are shown in Fig.\ \ref{fig:md_diseng}. For temperatures other than 20K, the total energy distributions exhibit several peaks. The minor peaks in Fig.\ \ref{fig:md_diseng} correspond with the annealing process of MD runs. The major peaks (labeled 3, 2 and 1 in Fig.\ \ref{fig:md_diseng}) are derived from different puckering configurations sampled in the process of equilibration to constant T. Correspondingly as shown in Fig.\ \ref{fig:md_func}, the fluctuations of autocorrelation functions (after thermal equilibrium is reached) reach an asymptotic state after $6.0 ps$. Each of the three peaks in Fig.\ \ref{fig:md_diseng} demonstrates a basin on the PEL of $\alpha$-G. These nearly degenerate quenched equilibrium states are trapped in distinct basins on the PEL, and quenched minima within one basin form a continuum metastable state around inherent structures. Details of variations in bond angles, bond lengths and total energies of these metastable states are shown in Table \ref{tab:info_para}. The total energy variation between the basins corresponding to the major three peaks are averaged as $1.405\times10^{-3} eV$, which is one order of magnitude higher than the energy of fluctuations within a basin. In spite of their different energy scales, these quenched configurations belonged to distinct basins share identical local bonding. The only difference is that they pucker in distinct ways, as shown in Fig.\ \ref{fig:eng_side}.

\begin{figure}[htb]
\includegraphics[scale=0.15]{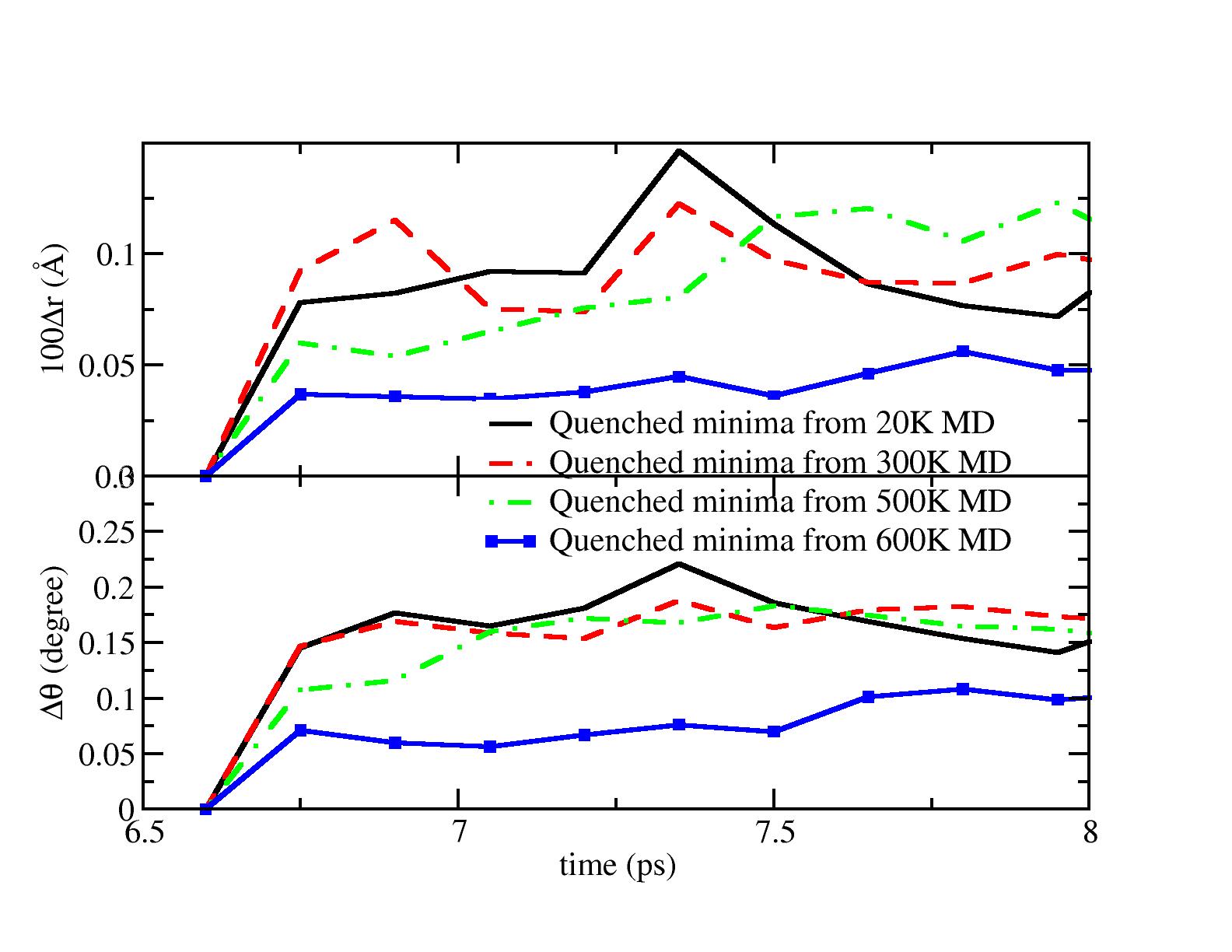}
\caption{Color. Time variation of two autocorrelation functions for $\alpha$-Si. This figure shows autocorrelation functions of $\Delta \theta(t_1,t_2)$ and $100\Delta r(t_1,t_2)$ for $t_1=6.6 ps$ and $t_2$ varying from 6.6 to 8.0 ps. The temperatures are 20K, 300K, 500K and 600K. The results are similar to \cite{md_drabold}.}
\label{fig:md_sifunc}
\end{figure}

For comparison, we repeat parallel calculations using 64 $\alpha$-Si model quenched from MD runs at 20K, 300K, 500K and 600K. The variations of autocorrelation functions are shown in Fig.\ \ref{fig:md_sifunc}. The results are in agreement with Fedders and Drabold\cite{md_drabold}. We see for $\alpha$-Si systems, there exists one general basin on the PEL (for a particular network connectivity), and the paths lowering the total energy on the PEL will eventually go into this basin, leading to inherent structures with minor changes in bond angles and lengths, and analogous energy scale.

Comparison between results of $\alpha$-G and $\alpha$-Si suggests that PEL of 3D system ($\alpha$-Si) is smooth and inherent structures are contained in one general basin. $\alpha$-G is similar within one puckered state.

\begin{figure}[htb]
\includegraphics[scale=0.3]{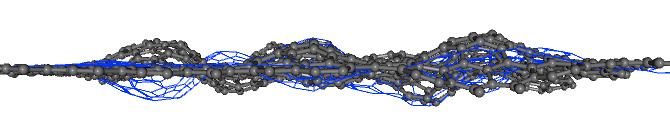}
\caption{Color. Side view of pucker-up and -down 800 a-g models. Gray balls and sticks illustrate pucker-up model, and pucker-down supercell is represented by blue lines.}
\label{fig:pucker_side}
\end{figure}

For $\alpha$-G, an MD run at higher temperature (annealing) can overcome the energy barrier between basins and reach an inherent structure with lower energy. Also by investigating into the correlation between topology and energy scale of these quenched supercells, it is revealed that lower total energy (stabler state) is related with more puckered configurations with small variation in bond lengths and angles from the original flat 800-atom model.

\subsection{Classical Normal Modes}
\label{sec:modes}

\begin{figure}
\includegraphics[scale=0.15]{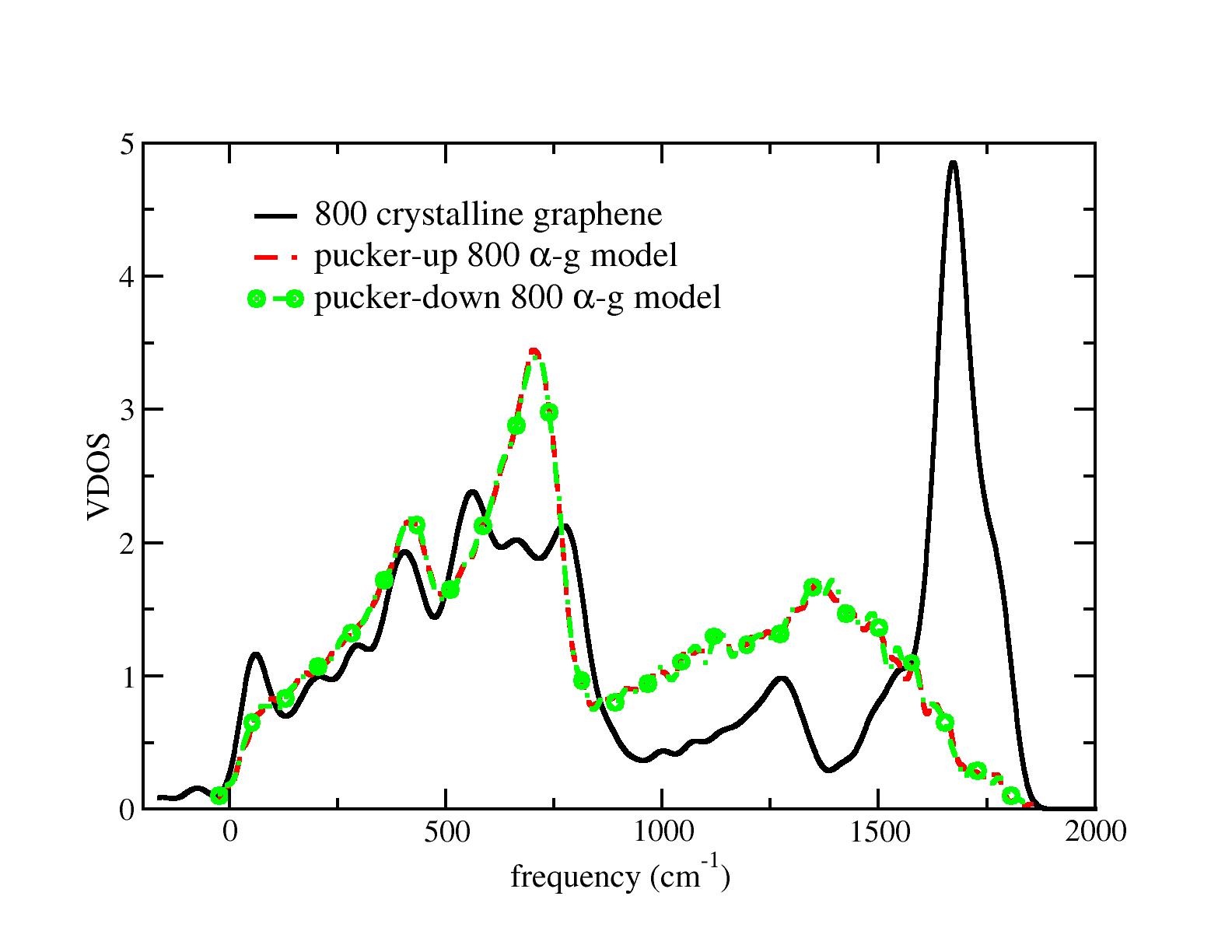}
\caption{Color. Vibrational density of states (VDOS) of 800 crystalline graphene, pucker-up and -down $\alpha$-g models. Note the distractive feature at $\omega\cong1375cm^{-1}$ for $\alpha$-G.}
\label{fig:com_vdos}
\end{figure}

\begin{figure*}[htb]
	\centering
	\subfloat[Imaginary-frequency mode in flat 800 $\alpha$-g model with frequency $\omega=283.843i cm^{-1}$ (referred as im-mode1).] {%
		\includegraphics*[width=0.45\textwidth]{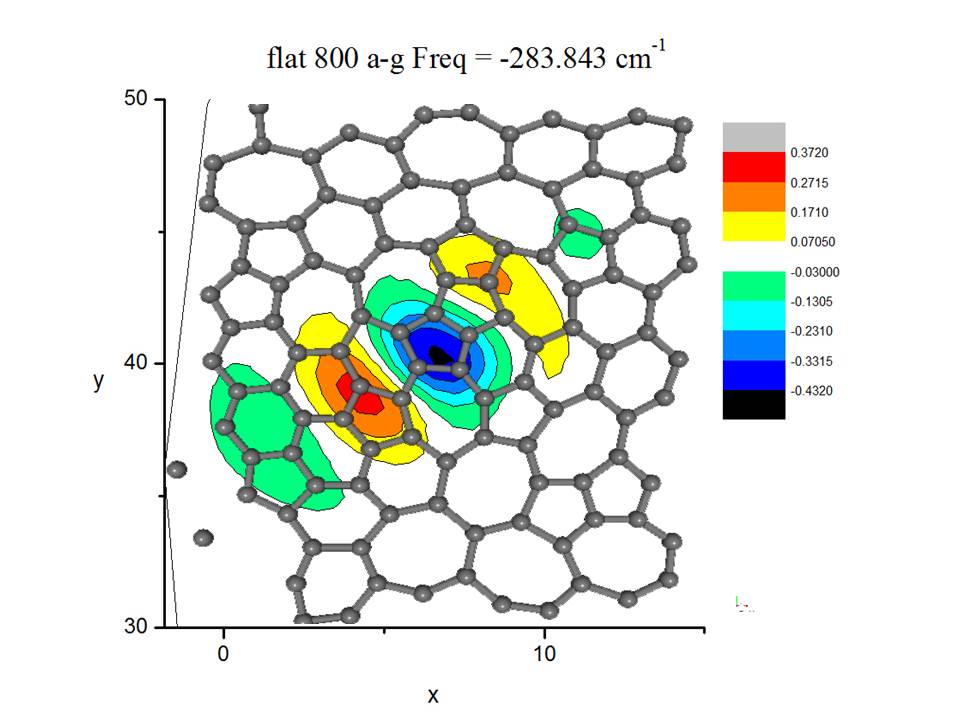}
		\label{fig:flat_neg1}
}
\hspace{0.5cm}
	\subfloat[Imaginary-frequency mode in flat 800 $\alpha$-g model with frequency $\omega=272.661i cm^{-1}$ (referred as im-mode2).] {%
		\includegraphics*[width=0.45\textwidth]{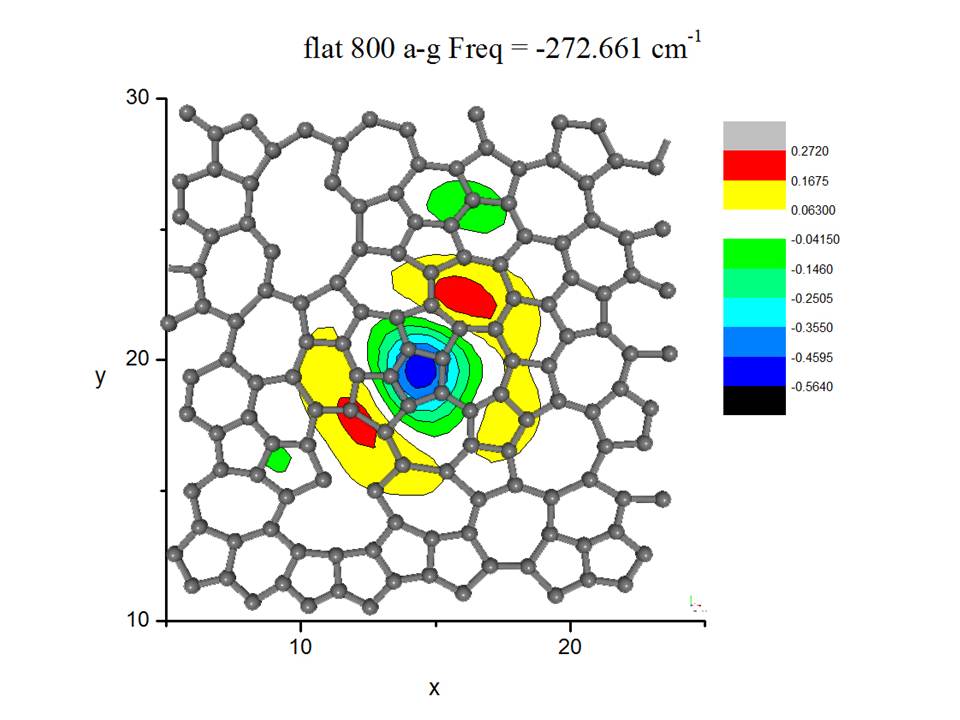}
		\label{fig:flat_neg2}
}\\
	\subfloat[Side view of the region where im-mode1 is originally localized.]{%
		\includegraphics*[width=0.45\textwidth]{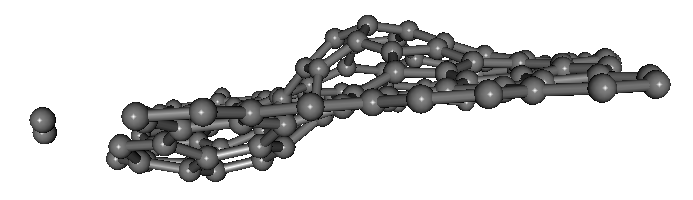}
		\label{fig:flat_rip1}
}
\hspace{0.5cm}
	\subfloat[Side view of the region where im-mode2 is originally localized.]{%
		\includegraphics*[width=0.45\textwidth]{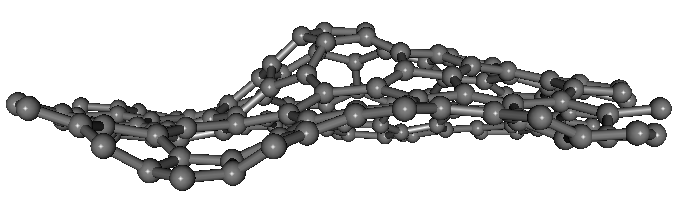}
		\label{fig:flat_rip2}
}
	\caption{Color. Two examples of imaginary-frequency modes in flat 800 $\alpha$-g model. The contour plot represents the component of eigenvector along the direction transverse to the plane.}
	\label{fig:flat_negmode}
\end{figure*}

\begin{figure}
\includegraphics[scale=0.2]{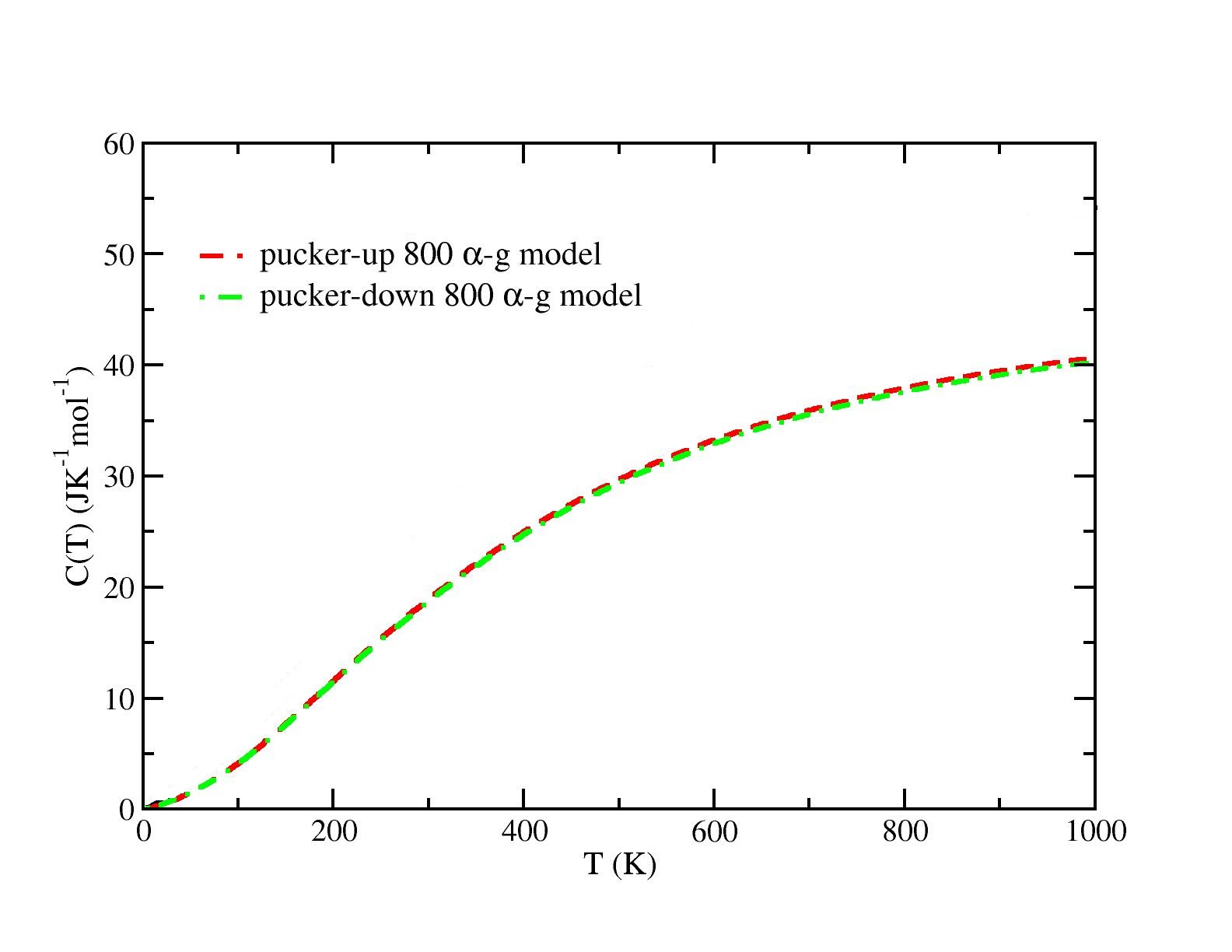}
\caption{Color. Temperature dependence of C(T) of pucker-up and -down 800 $\alpha$-g models.}
\label{fig:heat_temp}
\end{figure}

\begin{figure}[htb]
	\centering
	\subfloat[Low-frequency mode in pucker-down 800 $\alpha$-g model with frequency $\omega = 19.167 cm^{-1}$.]{%
	\includegraphics*[scale=0.32]{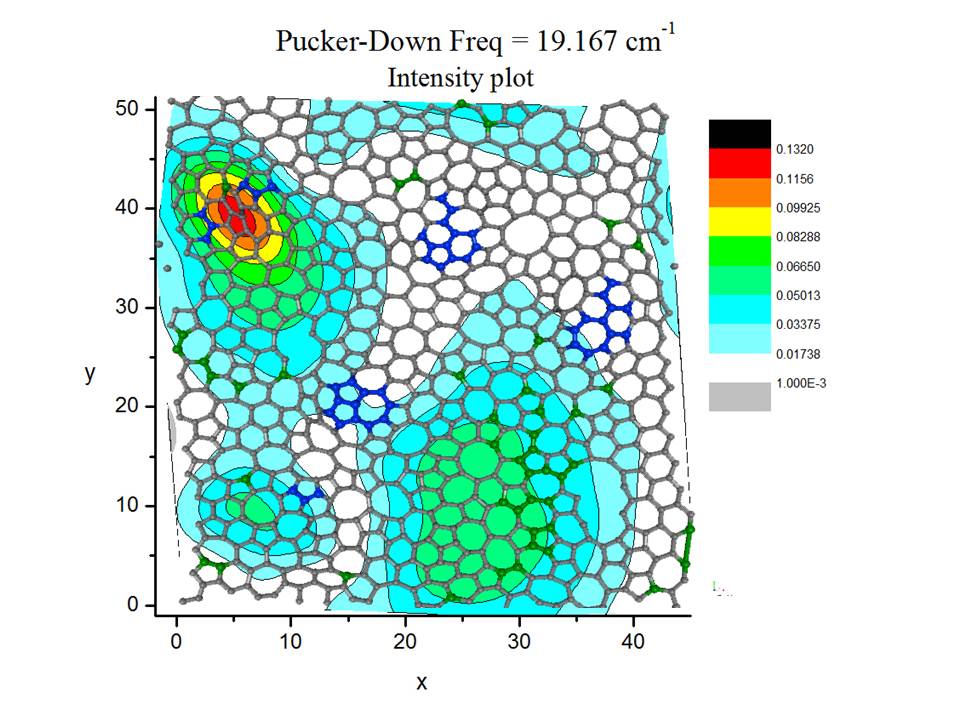}
	\label{fig:down_low}
}\\
	\subfloat[Low frequency mode in pucker-up 800 $\alpha$-g model with frequency $\omega = 17.318 cm^{-1}$.]{%
	\includegraphics*[scale=0.32]{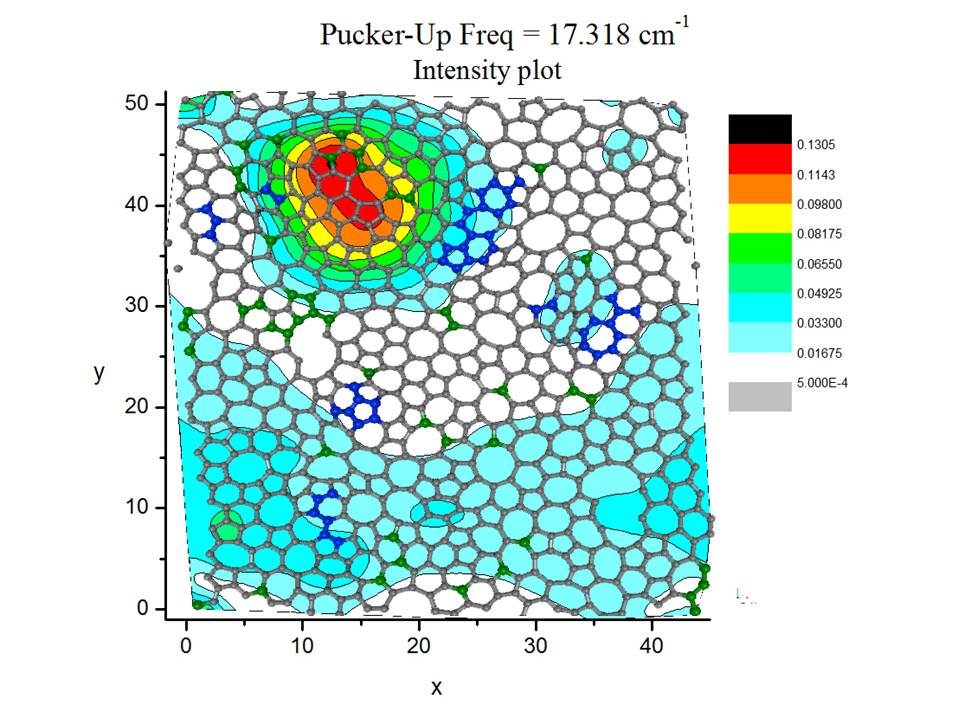}
	\label{fig:up_low}
}
\caption{Color. Examples of low-frequency modes in pucker-down and -up 800 $\alpha$-g models. The contour plots represent the intensity of eigenvectors on each atom. The blue atoms illustrate the ``puckering-most'' atoms, and the green atoms represent ``flat'' atoms.}
\label{fig:lowfreq_mode}
\end{figure}

To investigate the vibrational modes in these supercells, we perform calculations of dynamical matrix eigenvalues and eigenvectors of original flat 800 $\alpha$-g model, and two quenched configurations with certain region puckering along opposite direction, designated ``pucker-up'' and ``pucker-down'' models, as shown in Fig.\ \ref{fig:pucker_side}. The dynamical matrix was constructed by moving each atom along six directions by 0.04 Bohr. We also perform the phonon calculation for a 800-atom crystalline graphene model. The vibrational density of states (VDOS) of 800 crystalline model, pucker-up and -down 800 $\alpha$-g models are shown in Fig.\ \ref{fig:com_vdos}. The VDOS result of crystalline graphene shows good agreement with published calculation\cite{publish_vdos}. In Fig.\ \ref{fig:com_vdos} at a frequency near 1375 cm$^{-1}$, the spectrum of crystalline graphene has a minimum. In contrast the spectrum of two puckered supercells achieve a local maximum. Thus Raman scattering experiments are expected to provide a way to distinguish crystalline and amorphous graphene. There is no difference in the spectrum between pucker-up and -down 800 $\alpha$-g models. 

In the original flat 800-atom $\alpha$-g model, the eigenvectors with imaginary eigenvalues have large components along the normal direction of the graphene plane (at least four orders of magnitude higher than x and y components). These imaginary-frequency modes are localized on pentagons in the network: two example are shown in Fig.\ \ref{fig:flat_negmode}. As shown in Fig.\ \ref{fig:flat_rip1} and \ref{fig:flat_rip2}, these imaginary-frequency modes are localized near structures that lead to puckering. As shown in \cite{grap_pucker}, it is pentagons that lead to puckering and symmetry breaking. Thus these imaginary-frequency modes are an indicator of the instability of the flat 800 $\alpha$-g model.

In the puckered models, we observe modes with a low frequency, around 14-20 cm$^{-1}$, reminiscent of ``floppy modes'' proposed by Phillips and Thorpe\cite{floppy_phillips}\cite{floppy_thorpe}. The structures of these low-frequency modes are quite complex, as shown in Fig.\ \ref{fig:lowfreq_mode}. These modes are rather extended, and have significant weight on pentagonal puckered regions and large rings, analogous to what Fedders and Drabold have seen in $\alpha$-Si:H\cite{md_drabold}. The observed energy scale of these low-frequency modes is around a few $meV$, almost half of the lowest frequency of the acoustic phonon modes in a crystalline graphene with same size. As proposed in the theory of ``two-level systems'', there exists a vast distribution of low-energy excitations, caused by tunneling of atoms between nearly degenerate equilibrium states\cite{tls_anderson}\cite{tls_phillips}. Goldstein pointed out that the dynamics could be separated into two categories: vibrational motion about a minimum on PEL and transitions between minima\cite{goldstein}. Then these low-frequency modes might be triggered by transitions between degenerate minima within one basin on the PEL. As shown in Sec.\ \ref{sec:confor_fluc}, the energy variations between minima within one basin is in the order of $10^{-4}eV$, and the energy difference between basins is in the order of $10^{-3}eV$. The energy scales of these low-frequency modes ($\sim meV$) are sufficient to drive conformational fluctuations, but not high enough to overcome the energy barrier between different basins (quenched states) on the PEL.

\begin{figure}[htb]
	\centering
	\subfloat[High-frequency mode in pucker-down 800 $\alpha$-g model with frequency $\omega = 1642.073 cm^{-1}$.]{%
	\includegraphics*[scale=0.32]{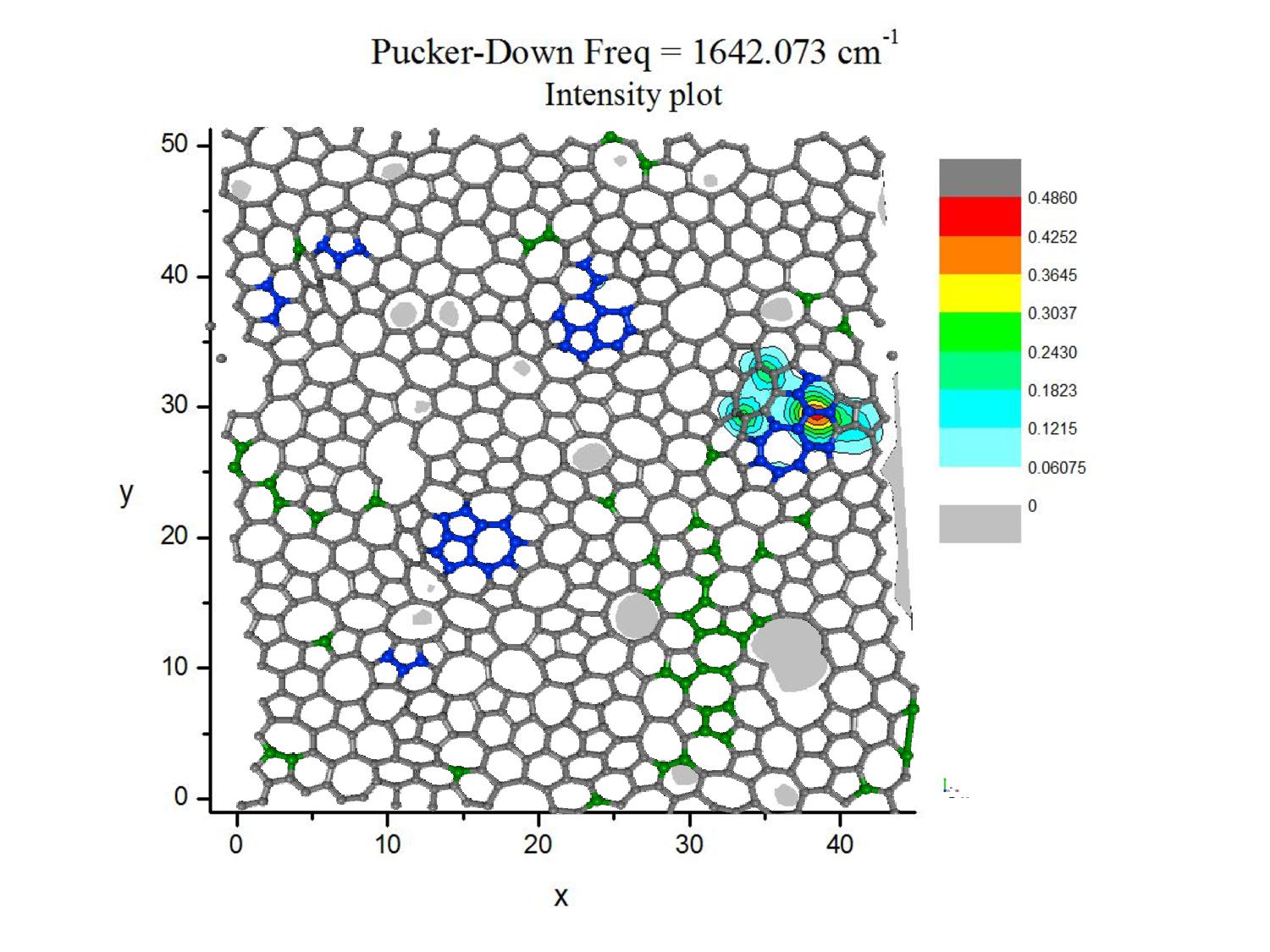}
	\label{fig:down_high}
}\\
	\subfloat[High-frequency mode in pucker-up 800 $\alpha$-g model with frequency $\omega = 1629.252 cm^{-1}$.]{%
	\includegraphics*[scale=0.32]{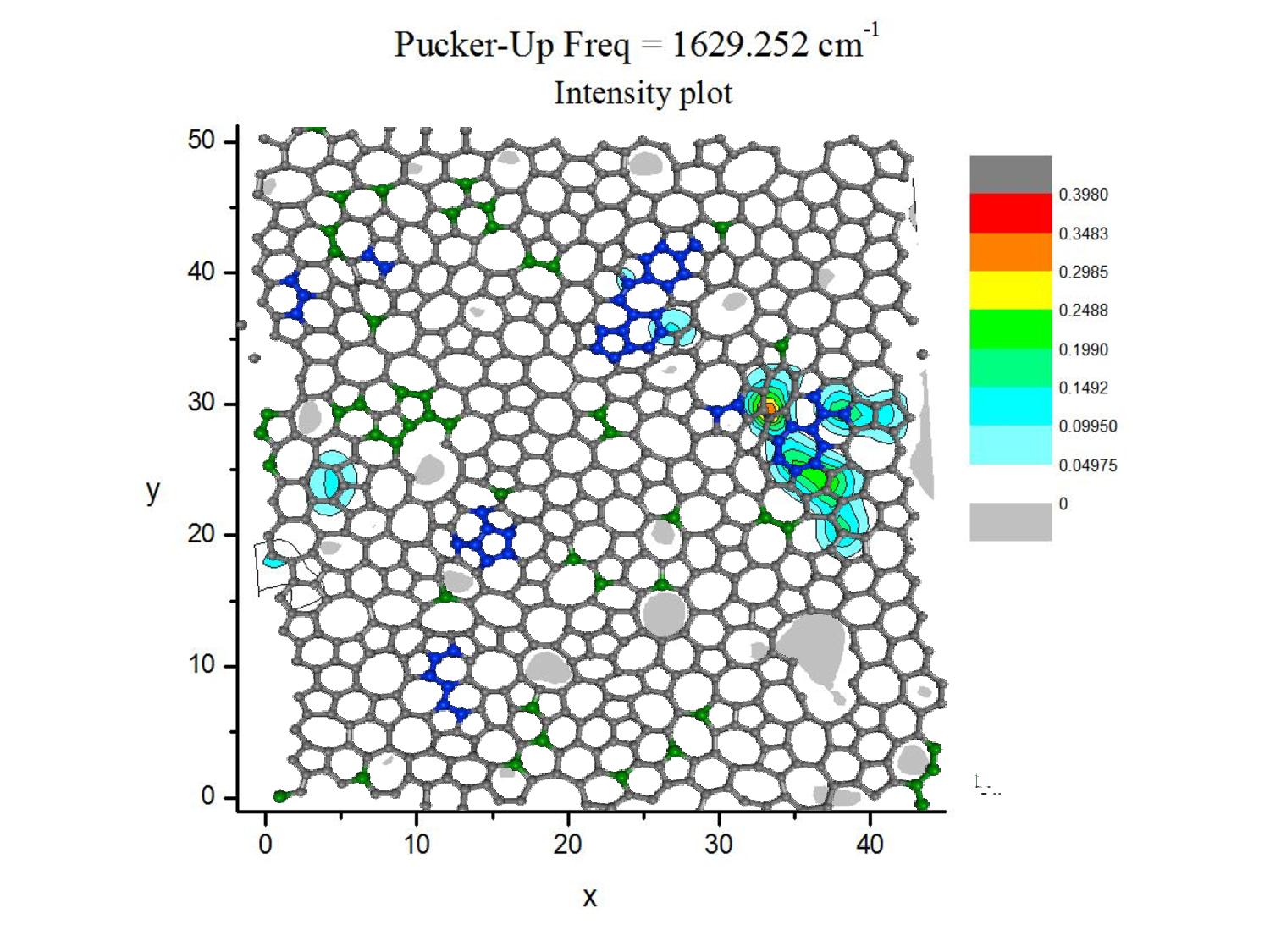}
	\label{fig:up_high}
}
\caption{Color. Examples of localized high-frequency modes in pucker-down and -up 800 $\alpha$-g models. The contour plots represent the intensity of eigenvectors on each atom. The blue atoms illustrate the ``puckering-most'' atoms, and the green atoms represent ``flat'' atoms.}
\label{fig:highfreq_mode}
\end{figure}

In the high-frequency domain, there also exist highly localized high-frequency modes in the puckered configurations. These modes are triggered by the pentagonal defects and are highly localized, as shown in Fig.\ \ref{fig:highfreq_mode}. This result is consistent with Biswas et al.\ \cite{bis_highfreq} and Fedders et al.\ \cite{fed_highfreq}, who have shown strain and topological defects are active at highest frequencies.

We also compute the specific heat $C(T)$ using VDOS information\cite{prb_heat}:
\begin{equation}
\label{eq:spec_heat}
C(T)=3R\int_0^{E_{max}} (\frac{E}{k_BT})^2\frac{e^{E/k_BT}}{(e^{E/k_BT}-1)^2}g(E)dE
\end{equation}
where g(E) is normalized VDOS. For room temperature (300K), specific heat of flat, pucker-up and -down 800 $\alpha$-g models are $25.151$, $18.879$ and $18.702 JK^{-1}mol^{-1}$ respectively. The temperature dependence of C(T) is shown in Fig.\ \ref{fig:heat_temp}. This is presumably an academic result as it is currently hard to imagine an experiment for $C(T)$ for this 2D system.

\section{Conclusion}
\label{sec:conclusion}

In conclusion, we have found that $\alpha$-G has a rich and interesting energy landscape. We observe distinct energy scale of basins ($\sim 10meV$) associated with different puckered configurations and then within such a configuration, an ambiguous energy minimum with a continuum of bond angles and bond lengths with energy scale ($\sim$ few$\mu eV$) and a nearly flat PEL. Within a given puckered configuration, this continuum is much like what was seen for $\alpha$-Si:H in 1996\cite{md_drabold}.

Vibrational calculations reveal the existence of localized imaginary-frequency modes in flat 800 $\alpha$-g model. These modes are localized on pentagons and play the key role in losing planar symmetry and forming pentagonal puckering structures. We find delocalized low-frequency phonon modes, similar to floppy modes, which have substantial weight on defects and share the same energy scale as the energy difference between adjacent basins on the PEL. Thus these low-frequency modes are triggered by the transition between adjacent energy minima. Some high-frequency modes are detected and highly localized on puckered regions and large rings.

\begin{acknowledgement}
Here we want to strongly acknowledge Dr.\ M.\ F.\ Thorpe at Arizona State University and his former student Harry He, who have prepared the 800-atom $\alpha$-G model. This work is supported by NSF Grant No.\ DMR 09-03225.
\end{acknowledgement}

\end{document}